\documentclass[final,3p,times,twocolumn]{elsarticle}

\usepackage{graphics}
\usepackage{epstopdf}

\usepackage{amssymb}

\usepackage{ulem}
\usepackage{color}

\newcommand{\RE}{\mathrm{Re}}
\newcommand{\IM}{\mathrm{Im}}

\begin{document}

\begin{frontmatter}

\title{Quantum mechanical model for J$/\psi$ suppression in the LHC era}

%

\author[address1]{C. Pe\~{n}a}
\ead{pena@ift.uni.wroc.pl}

\author[address1,address2,address3]{D. Blaschke}
\ead{blaschke@ift.uni.wroc.pl}

\address[address1]{Institute for Theoretical Physics, University of Wroc{\l}aw,
pl. M. Borna 9, 50-204 Wroc{\l}aw, Poland}
\address[address2]{Bogoliubov Laboratory for Theoretical Physics, JINR Dubna,
Joliot-Curie str. 6, 161980 Dubna, Russia}
\address[address3]{Fakult\"at f\"ur Physik, Universit\"at Bielefeld, 
Universit\"atsstra{\ss}e 25, 33615 Bielefeld, Germany}


\begin{abstract}
We discuss the interplay of  screening, absorption and regeneration effects, 
on the quantum mechanical evolution of quarkonia states, within a 
time-dependent harmonic oscillator (THO) model with complex oscillator 
strength. 
We compare the results with data for $R_{\rm AA}/R_{\rm AA}$(CNM) from CERN 
and RHIC experiments.
In the absence of a measurement of cold nuclear matter (CNM) effects at LHC we 
estimate their role and interpret the recent data from the ALICE experiment. 
We also discuss the temperature dependence of the real and imaginary parts of 
the oscillator frequency which stand for screening and absorption/regeneration,
respectively. 
We point out that a structure in the J$/\psi$ suppression pattern for In-In
collisions at SPS is possibly related to the recently found $X(3872)$
state in the charmonium spectrum. 
Theoretical support for this hypothesis comes from the cluster expansion of 
the plasma Hamiltonian for heavy quarkonia in a strongly correlated medium.
\end{abstract}


\end{frontmatter}

\section{Introduction}
\label{Introduction}

The J/$\psi$ suppression at CERN and RHIC has been quantified by measurements 
of the nuclear modification factor  $R_{AA}$ \cite{Ferreiro:2009ur}. 
This set of experimental data is commonly normalized to the baseline of 
cold nuclear matter (CNM) effects $R_{\rm AA}(\textrm{CNM})$. 
The results are scaled with the multiplicity of charged particles 
$\frac{dN_{\rm ch}}{d\eta}\big|_{\eta=0}$ 
\cite{Arnaldi:2012qa, Arnaldi:2009, Arnaldi:2007zz, Adler:2004zn}.

A key problem to be solved in heavy-ion collisions is that the usage of heavy 
quarkonia states as a probe for the diagnostics of the quark-gluon plasma 
(QGP) requires the knowledge of the baseline for their production and 
evolution characteristics in situations when a QGP is absent. 
For a review see, e.g., \cite{Rapp:2008tf} and references therein.
Current issues in the experiment-theory dialogue are summarized, e.g., in
\cite{Brambilla:2010cs}. 
Aspects of quarkonium production at LHC are discussed, e.g., in
\cite{Bedjidian:2003gd,Lansberg:2008zm} and new information for its 
$R_{\rm AA}(\textrm{CNM})$ baseline are given in \cite{vogt:2012nf}.

The suppression of charmonium (more general, heavy quarkonium suppression), 
due to color screening, was suggested to be a signal of QGP formation 
\cite{Matsui:1986dk,Ding:2010yz}. 
The works \cite{Jankowski:2009kr,Blaschke2011137,Blaschke:2005jg} have studied 
the dissociation of quarkonium (charmonium and bottomonium)  by screening in a 
hot plasma. 
The binding energy of the heavy quark-antiquark pair in the quarkonium state 
vanishes at the respective Mott temperature where the bound state merges the 
continuum of scattering states \cite{Burau:2000pn, Blaschke:2002ww}.
Below the Mott temperatures, the binding energies are already sufficiently 
lowered so that collisions with particles from the medium may have sufficient 
thermal energy to overcome the threshold for impact dissociation of quarkonium 
\cite{Ropke:1988bx,Ropke:1988zz,Blaschke:2003ji,Blaschke:2004dv}. 
Both effects tend to wash out the pattern of sequential suppression for heavy 
quarkonia states expected from the ``classical'' picture of the Mott effect
\cite{Digal:2001ue}. 
The description of quarkonia states in a hot QGP medium in the vicinity of the
critical temperature should therefore treat bound and scattering states on an
equal footing. 
This is appropriately achieved within a thermodynamical T-matrix approach,
which has been developed to address the spectral properties of quarkonia 
\cite{Cabrera:2006wh} as well as open flavor meson states
\cite{vanHees:2007me,vanHees:2008gj,Riek:2010fk}.  
It is also worth noting that inelastic collisions are responsible for 
absorption of J/$\psi$ in a hadron gas  of light $\pi$ and $\rho$ mesons
since the corresponding cross sections can be sufficienly large
\cite{Blaschke:2004dv, Wong:1999zb, Ivanov:2003ge, Bourque:2008es, Blaschke:2008mu}. 
Once the partial densities of open charm mesons becomes nonnegligible, also
the reverse processes of charmonium regeneration in channels like, e.g., 
$D + \bar{D}\to$ J/$\psi$  + $\rho$ will occur.  
J/$\psi$ regeneration shall give important contributions to charmonium 
production in heavy-ion collisions already at RHIC 
\cite{Ko:1998fs,BraunMunzinger:2000dv,BraunMunzinger:2000px,Thews:2000rj,Grandchamp:2001pf,Grandchamp:2002wp}
and the more at the LHC 
\cite{Andronic:2010dt,Andronic:2011yq}.
The situation is well summarized in two recent reviews 
\cite{Kluberg:2009wc,BraunMunzinger:2009ih}.

Another very interesting alternative is the channel 
$D + \bar{D}^*\to$ J/$\psi$  + $\rho$ \cite{Blaschke2011137} where as a 
``molecular'' bound state of $D$ and $\bar{D}^*$ with binding energy below 
$1$ MeV the exotic state X(3872) is discussed 
\cite{Brambilla:2010cs, Burns:2010qq,Carlosx:2011cp} which was recently
discovered by BaBar \cite{Aubert:2008gu} and subsequently confirmed by other 
experiments.

It is a main assumption of the present work that the reactions between hadrons
which were discussed above can take place in the strongly coupled QGP (sQGP)
phase as well, but there between the ``pre-formed hadrons'', i.e. resonances 
in the strongly coupled quark-antiquark plasma. 
Evidence for the existence of such resonances in the sQGP comes from lattice 
QCD (LQCD) analyses of correlation functions (for recent references see, e.g., 
\cite{Petreczky:2003iz,Ding:2012iy})
and is expected in accordance with the Mott mechanism 
\cite{Burau:2000pn,Blaschke:2002ww,Blaschke:2003ji,Blaschke:2004dv}.
The strong increase of rate coefficients for flavor kinetic processes between
hadronic resonances in the vicinity of the hadronization transition 
\cite{BraunMunzinger:2003zz}
has been a main ingredient to a recent model for chemical freezeout in 
heavy-ion collisions  \cite{Blaschke:2011ry,Blaschke:2011hm}.

In the present work we consider the quantum evolution of $Q\bar{Q}$  
correlations in the sQGP phase which get projected onto the 
well-known hadronic basis states of the vacuum charmonium spectrum  
at the heavy quarkonia freeze-out temperature $T_f$, which is above the
critical temperature $T_c$ of the QCD phase transition. 
Actually, we shall neglect here reactions of hadronic states containing
heavy quarks for $T<T_f$. 
In particular, we do not consider any rearrangement collisions in the hadron 
gas phase after hadronization.   

It has been shown that as long as medium effects can be embodied in a gaussian
action, the $Q\bar{Q}$ propagator obeys a closed temporal evolution equation
whose large-time behavior is governed by an effective potential. 
The latter, besides screening, displays also an imaginary part related to 
collisions \cite{Beraudo:2007ky,Laine:2007qy}. 
The determination of this imaginary part also provides insight into how the 
anisotropy of the medium relates with the width of the states 
\cite{Dumitru:2009, Strickland:2011s}.  
The use of complex potentials for describing data of charmonium suppression  
in $pp$ and $pA$ collisions has been documented in 
\cite{Quack:1991yf,Cugnnon:1993yf,Koudela:2003yd,Martins:1994hd}. 
An application of a time-dependent harmonic oscillator (THO) model with a 
complex oscillator frequency to experimental data of charmonium production in 
$AA$ collisions has recently been given \cite{Blaschke2011137,Carlos:2011cp}. 
This application predicted the formation of resonances nearby the onset of 
charmonium suppression (Mott transition) \cite{Carlosthesis:2012}. 
This model exhibits confinement in the sense that the potential grows 
quadratically with the separation distance of the  $Q\bar{Q}$ pair. 

Here we employ the THO model for studying the role of static screening and 
absorption/regeneration kinetics in the quarkonium time evolution, thus 
generalizing the Matsui formula \cite{Matsui:1989ig} to the case of a complex 
$Q\bar{Q}$ potential. 
The quantum mechanical evolution, as a function of the time $t$ is 
encrypted in the evolution amplitude $U(\vec{r},\vec{r}_0,t)$ for the 
initial $Q\bar{Q}$ state $\phi_{Q\bar{Q}}$. 
We have examined two alternatives for the initial state: 
a delta function \cite{Blaschke2011137,Carlos:2011cp} and 
a Gaussian \cite{Carlosthesis:2012}. 
Other forms are also possible like the Bessel function used in a 
description of the photoproduction of charmonia \cite{Kopeliovich:1991pu}.
The time dependence originates from the temperature evolution  of the medium 
$T(\tau)$ which is assumed to obey total entropy conservation. 
In order to account for effects of the hot medium produced in $\rm AA$ 
collisions, 
the temperature dependence of the complex oscillator frequency $\omega^2(T)$ 
stands for screening, absorption and regeneration in a unified approach. 
By taking different ans\"{a}tze for $\omega^2(T)$, we show that the THO model 
can describe the behavior of the measured J/$\psi$ yield over the expected 
CNM effects $R_{\rm AA}/R_{\rm AA}$(CNM), as a function of the multiplicity of 
charged particles. 
Our method can be applied to more realistic cases, even for a  $Q\bar{Q}$ 
Hamiltonians containing a complex confinement potential as the ones given in 
\cite{Dumitru:2009, Strickland:2011s} which are in good agreement with LQCD.

We pay special attention to the In-In data \cite{Arnaldi:2007zz}
taken at CERN SPS since they have the highest precision. 
Interestingly, they exhibit a peculiar shape (``wiggle") 
\cite{Blaschke2011137,Carlosthesis:2012}. 
The wiggle turns out to be important since it appears around the estimated 
temperature for dissociation  of charmonium ($\sim 1.25~T_c$ with 
$T_c = 0.154$ GeV \cite{Borsanyi:2010cj}), which can be defined as 
the temperature for its Mott transition.
Our novel results obtained with the complex THO model suggest that such 
peculiar behavior can be mapped into $\IM[\omega^2(T)]$, the 
imaginary part of the  $Q\bar{Q}$ potential. 
We examine whether the production of the X(3872) resonance can contribute to 
the in-medium  $Q\bar{Q}$ hamiltonian.

\section{Suppression factor for quarkonium}

The survival probability (suppression factor) for J/$\psi$ is defined in
\cite{Matsui:1989ig}
\begin{eqnarray}
\label{spsi}
S_{\psi_n}(t)&=&
\Biggl|\frac{\int\;d^{3}r\;\psi_n^{*}(\vec{r})\;\phi_{Q\bar{Q}}(\vec{r},t)}{\int\;d^{3}r\;\psi_n^{*}(\vec{r})\;\phi_{Q\bar{Q}}(\vec{r},0)} \Biggl|^2,
\end{eqnarray}
This expression was published in bra-ket representation by Cugnon et al. 
\cite{Cugnnon:1993yf}.  
This is a generalization of the Matsui approach to an arbitrary initial state 
\cite{Carlosthesis:2012}. 
The intermediate $Q\bar{Q}$ state is given by
\begin{equation}
\phi_{Q\bar{Q}}(\vec{r},t) = 
\int\;d^{3}r_0\;\phi_{Q\bar{Q}}(\vec{r}_0,0)\;U(\vec{r},\vec{r}_0,t)~,
\end{equation}
where $\phi_{Q\bar{Q}}(\vec{r}_0,0)$ represents the initial $Q\bar{Q}$ state 
at the position $\vec{r}_0$ and time $t=0$.
In the following we shall assume that the quantum number $n$, on the 
quarkonium state $\psi_n$, stands for the principal, angular and azimuthal 
quantum numbers. 
The suppression factor can be generalized to the case of nonsingular complex 
potentials and be applied to the QGP diagnostics in collisions of heavy ions 
with mass number A when identified with the experimentally determined quantity
\begin{equation}
\label{suppression}
\frac{R_{\rm AA}}{R_{\rm AA}({\rm CNM})}=:S_{\psi_n}(t)
\end{equation}
where $R_{\rm AA}$(CNM) accounts for the CNM effects from charmonium 
absorption in cold nuclear matter and modification of charm production by 
shadowing/antishadowing of gluon distribution functions in the center of mass 
of the colliding nuclei. 
Both effects are accessible by analysis of $pA$ collision experiments, see
\cite{Ferreiro:2009ur,Rapp:2008tf} and references therein.
We restrict our discussion here to ground state charmonium at rest in the QGP
medium ($p_T=y=0$) so that the discussion of Lorentz boost effects on the
formation process can be omitted.
We include, however, the feed-down from higher charmonia states which are 
discussed more in detail elsewhere \cite{Carlos:2011cp,Carlosthesis:2012}.

The next section is dedicated to the application of the generalized Matsui 
approach. 
To this end, we start calculating the evolution amplitude 
$U(\vec{r},\vec{r}_0,t)$. 
As an example we consider the THO Hamiltonian to model the $Q\bar{Q}$ 
interaction.

\section{Time-dependent harmonic oscillator model}

For our discussion of the quantum mechanical evolution of quarkonia in an
evolving QCD plasma state, we will employ here a generalization of the
harmonic oscillator model \cite{Matsui:1989ig} to a time-dependent one with
nonsingular complex squared oscillator frequency (THO model). 
The merit of such a model is its simplicity and transparency as well as its
tractability \cite{Blaschke2011137,Carlos:2011cp,Carlosthesis:2012}. 
Aspects of an optical potential for the propagation of charmonia through a 
medium have been already discussed, e.g., for cold nuclear matter in
\cite{Koudela:2003yd,Kopeliovich:2003cn} and for a quark-gluon plasma in
\cite{Cugnnon:1993yf,Cugnon:1993ye}.
We consider the time-dependent Hamiltonian for heavy quarkonia in the form
\begin{equation}
H(t)=\frac{p^2}{2\mu}+\frac{\mu}{2}\omega^2(t) r^2(t)~,
\label{hamilton}
\end{equation}
where $\mu=m_Q/2$ is the reduced mass and $m_Q$ the heavy quark mass.
The squared complex oscillator frequency $\omega^2(t)$ has an implicit time
dependence due to the
temperature evolution $T(\tau)$ of the system surrounding the evolving
heavy quarkonium state at time $\tau=\tau_0 + t$.
The quadratic dependence of the imaginary part of the (optical)
oscillator potential is motivated by the phenomenon of color transparency,
see also \cite{Blaschke:1992pw} and references therein.\\

The general classical trajectories for the Hamiltonian (\ref{hamilton}) can be 
found in \cite{Blaschke2011137,Carlos:2011cp,Carlosthesis:2012,Gjaja:1992,Ermakov:2008yd,polyanin:handbook}
and references therein. 
It is given by a linear combinations of the two solutions
\begin{eqnarray}
\label{e2}
r_{1,2}(t)=\rho(t)\exp(\pm i\phi(t))~,~~
\phi(t)=\int\frac{ dt^\prime}{\rho^2(t^\prime)}~.
\end{eqnarray}
The amplitude $\rho(t)$ fulfills the Ermakov equation
\cite{Carlosthesis:2012,Ermakov:2008yd,polyanin:handbook} 
\begin{eqnarray}\label{ermakov}
\label{e3}
\ddot{\rho}(t)+\omega^2(t)\ \rho(t) -\frac{1}{\rho^3(t)}=0~,
\end{eqnarray}
for which exact solutions exist.  
The initial and final conditions of the oscillator are defined as 
$\vec{r}(0)=\vec{r}_0$ and  $\vec{r}(t)=\vec{r}$,  respectively. 
This allows to evaluate the transition amplitude as a function of time using 
path integral methods \cite{Carlos:2011cp,Carlosthesis:2012,kleinert}
\begin{eqnarray}
\label{evolution}
U(\vec{r},\vec{r}_0,t)&=&
\left[\frac{\mu~\rho(t)~ \rho^{-1}(0) \dot{\phi}(t)}
{2\pi i \sin(\phi(t)-\phi(0))}\right]^{3/2}\;{\rm e}^{i S_{\rm cl}}~,
\end{eqnarray}
where the classical action is exactly calculated by 
$S_{\rm cl}=\frac{\mu}{2}~\Bigl(\vec{r}\cdot\dot{\vec{r}}
-\vec{r}_0\cdot\dot{\vec{r}_0}\Bigl)$ 
\cite{Carlosthesis:2012,kleinert}.
Additioanlly the initial contitions for the Ermakov equation (\ref{ermakov}) 
are defined as $\rho(0)=\frac{1}{\sqrt{\omega(0)}}$ , $\dot{\rho}(0)=0$.\\

\section{Applications to Heavy Ion Collisions}

\begin{figure}[htb]
      \centering
\includegraphics[width=0.9\columnwidth]{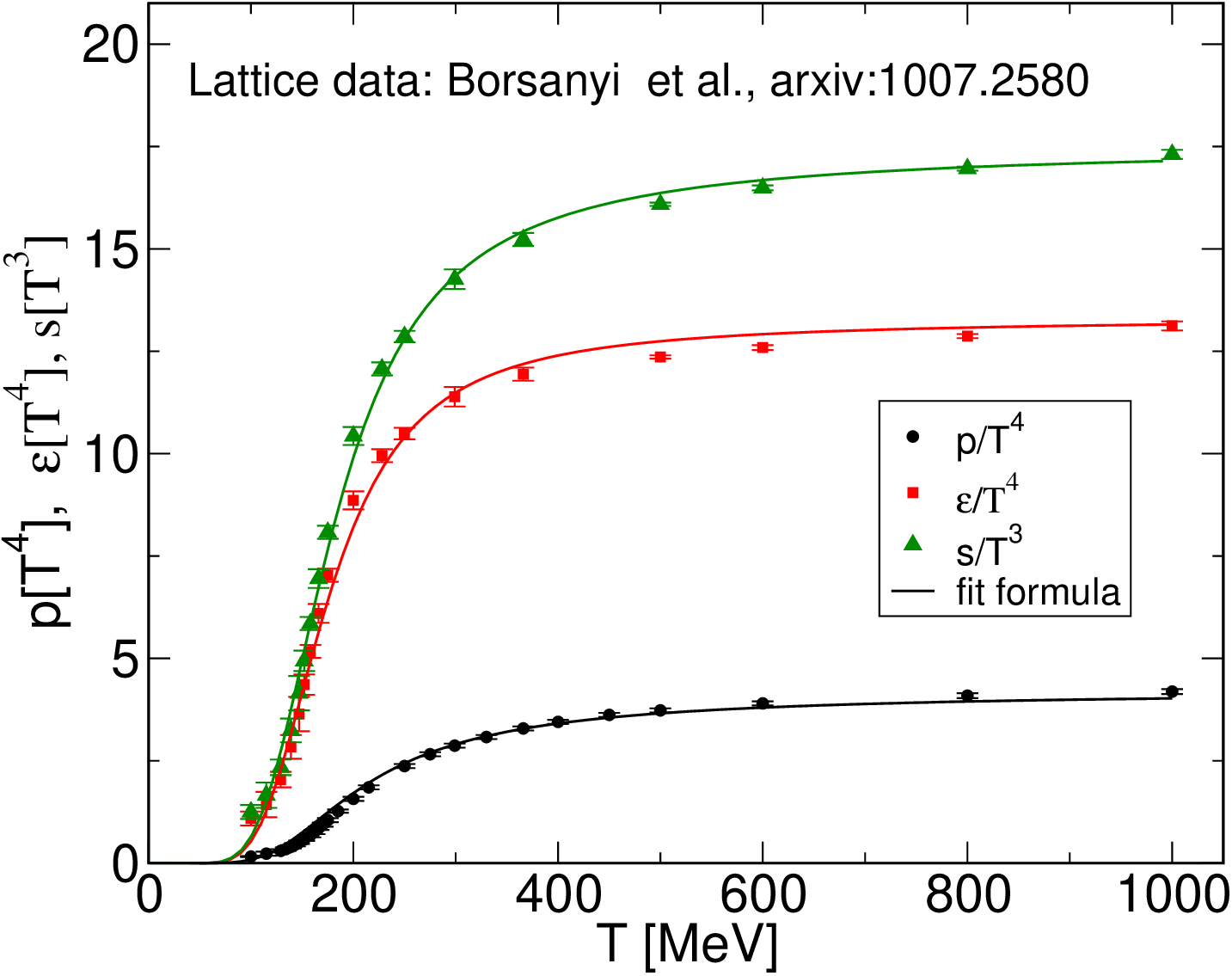}
\caption{Recent results (data points) for the equation of state from
lattice QCD \cite{Borsanyi:2010cj} compared to the fit formula (\ref{eos-fit})
employed here.}
	\label{fig:eos}
\end{figure}

In the following we show that the anomalous J/$\psi$ suppression in SPS, RHIC, 
and LHC experiments can be simultaneously described with the natural assumption
that above the critical temperature the relevant screening of the quarkonium 
interaction can be parameterized with a temperature dependent, complex 
oscillator strength.
The time evolution of the temperature itself will be given by longitudinally 
boost invariant (Bjorken scaling) hydrodynamic evolution of a fireball volume 
$V(\tau)$ under entropy conservation
\begin{equation}
\label{entropy}
s(T(\tau))V(\tau)={\rm const}~;~~V(\tau)=A_T \tau~.
\end{equation}
The temperature dependence of the entropy density $s(T)$ is taken
from recent lattice QCD simulations \cite{Borsanyi:2010cj} which are well
parameterized by the simple ansatz \cite{Blaschke2011137,Carlosthesis:2012}
\begin{equation}
\label{eos-fit}
s(T)=9.0~T^3~\left[1+\tanh\left(\frac{T-0.189}{0.534~T} \right) \right]~,
\end{equation}
see Fig.~\ref{fig:eos}. 

\begin{figure}[htb]
	\centering
	\includegraphics[width=0.9\columnwidth]{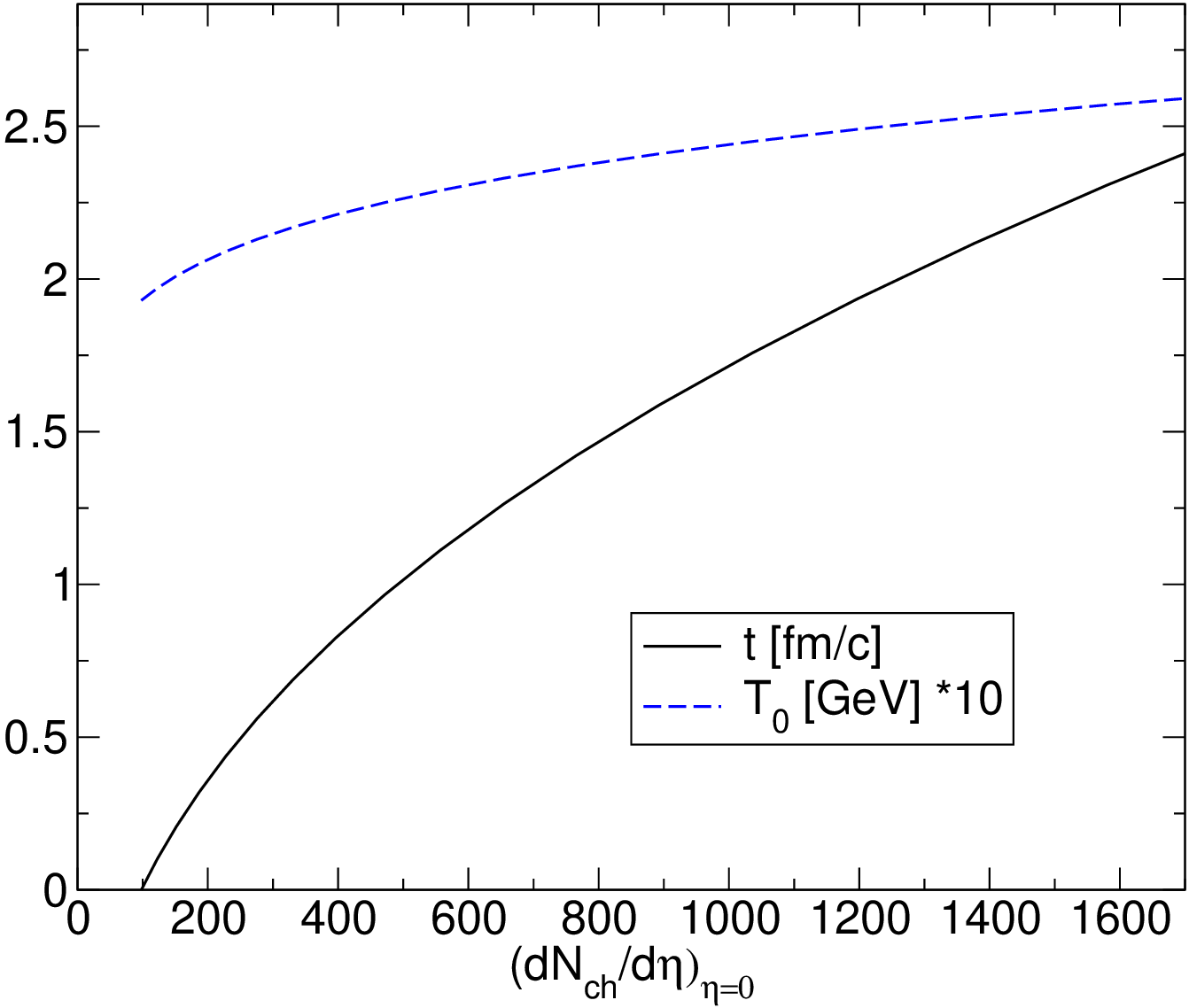}
	\caption{Initial temperature $T_0$ and time interval $t=\tau_f-\tau_0$
for the quantum evolution of the $Q\bar{Q}$ state from formation at $\tau_0$ to 
freeze-out at $\tau_f$ as a function of the measured charged particle yields 
per pseudorapitity interval $(dN_{ch}/d\eta)_{\eta=0}$.}
	\label{fig:hydro}
\end{figure}
The initial values of temperature correspond to initial entropy densities
given by the Bjorken formula \cite{Hatsuda}
\begin{equation}
\label{Bjorken-for}
s_0=\frac{3.6}{A_T \tau_0}
\left(\frac{\mathrm{d} N}{\mathrm{d} \mathrm{y}} \right )_{\mathrm{y}=0},
\end{equation}
where the transverse overlap area $A_T$ is taken in this work as 
\cite{Blaschke2011137,Carlosthesis:2012}
\begin{equation}
\label{AT}
A_T=(4.8\;\textrm{fm}^2)\left[\left(\frac{{\rm d} N_{\rm ch}}{{\rm d}{\eta}} 
\right)_{{\eta}=0} \right]^{0.57}.
\end{equation}
We found the parametrization by fitting the average values of the overlap area 
for PbPb. 
The latter was obtained from Monte Carlo calculations performed in 
Ref.~\cite{Adler:2004zn}. 
Additionally the transformation \cite{Adler:2004zn,Hatsuda}
{$
\left(\frac{{\rm d} N}{{\rm d}y}\right)_{y=0} = 1.04 \times \frac{3}{2} \times 
\left(\frac{{\rm d} N_{\rm ch}}{{\rm d}\eta}\right)_{\eta=0}$} 
has been used.

We are interested in the time  evolution of the quarkonium state in the 
cooling fireball medium described by the thermal history $T(\tau)$ starting 
from $\tau=\tau_0 = 1.0$  fm/c until the freeze-out of the effective THO 
potential parameters at $\tau_f=\tau_0 s(T_0)/s(T_f)$ with $T_f=T(\tau_f)=0.193$ GeV \cite{Carlosthesis:2012}.
In  Fig.~\ref{fig:hydro} we show the initial temperature $T_0(t)$ and 
the duration of the quarkonium evolution until its freeze-out 
$t=\tau_f-\tau_0$ as a function of the charged particle multiplicity per 
pseudorapidity interval $(dN_{\rm ch}/d\eta)_{\eta=0}$ characterizing the 
initial conditions of the fireball evolution\footnote{In using Eq.~(\ref{spsi})
for describing anomalous J/$\psi$ suppression due to a plasma medium 
switched on at a time $\tau_0$ after the creation of the $Q\bar{Q}$ pair 
implies that the evolution of the latter before $\tau_0$ is assumed to be
governed by a hermitian Hamiltonian which does not yet suffer a modification.}.

We consider the most suitable parametrization of the THO model with complex 
squared frequency given by\footnote{Note that in Ref.~\cite{Blaschke2011137} a 
different convention for the THO frequency has been used.}
$\omega^2(T)=\RE[\omega^2(T)]+i~\IM[\omega^2(T)]$. 
In this case screening, absorption and regeneration can be described 
simultaneously. 
Screening occurs when the real part of the squared frequency is 
$\RE[\omega^2(T)] < \omega_\psi^2$, where $\omega_\psi$ is the oscillator 
frequency for the J/$\psi$ in vacuum. 
The case of absorption appears when the imaginary part is 
$\IM[\omega^2(T)]<0$. 
The regeneration scenario 
\cite{Ko:1998fs,BraunMunzinger:2000dv,Thews:2000rj,Grandchamp:2001pf,Grandchamp:2002wp,Andronic:2010dt,Andronic:2011yq}
corresponds to $\IM[\omega^2(T)]>0$ in the THO model which stems from the 
coupling of the charmonium hamiltonian to open charm channels.
In addition we shall consider a gaussian wave function for initial $Q\bar{Q}$ 
state. Details are presented in \cite{Carlosthesis:2012}.

The ground state of quarkonium can be identified with the 1S state of the 
harmonic oscillator \cite{Carlos:2011cp,Matsui:1989ig} given by 
$\psi_{000}(\vec{r})=\psi_{000}(0)\exp\Bigl(\frac{-r^2}{2~r^2_\psi}\Bigl)$ 
with $ r_\psi=\sqrt{\frac{1}{\mu~\omega_\psi}}$. 
Calculations of the suppression factor with Dirac delta-shaped initial 
$Q\bar{Q}$ state were performed in \cite{Blaschke2011137} on the basis
of Ref.~\cite{Carlos:2011cp}.
In the following we use results for the J/$\psi$ survival probability 
obtained in \cite{Carlosthesis:2012} using a gaussian shaped initial state 
and discuss the relation between parametrizations of the temperature dependent 
complex squared oscillator strength to experimental data on anomalous J/$\psi$ 
suppression from SPS, RHIC and LHC.
In particular, we discuss how accounting or not for the ``wiggle'' in the 
In-In data at SPS when parameterizing $\omega^2(T)$  will change predictions 
of the survival probability in the LHC domain. 

\section{Anomalous suppression from SPS to LHC}

Anomalous suppression, the deviation from unity of experimental data for the 
J/$\psi$ production ratio normalized to the expectation accounting for CNM 
effects (\ref{suppression}), is considered as a key indicator for QGP 
formation in heavy-ion collisions.
This effect, first observed at CERN SPS for Pb-Pb collisions at $\sqrt{s}=17$
GeV, has been qualitatively confirmed by RHIC experiments with Au-Au
interactions at $\sqrt{s}=200$ GeV whereby three particularly interesting
observations were made 
\begin{itemize}
\item[(i)] the suppression is stronger at forward and backward rapidities
rather than at midrapidity where the particle densities are highest,
\item[(ii)] the onset of anomalous suppression and its dependence on centrality
scales with the charged particle density at midrapidity rather than with energy
density. 
\item[(iii)] the rather precise data of the NA60 collaboration for In-In
collisions show a dip in the centrality dependence of the anomalous 
J/$\psi$ suppression ratio \cite{Arnaldi:2009,Arnaldi:2007zz}. 
\end{itemize}
While (i) is caused mainly by antishadowing and to some extent by geometry
\cite{Prorok:2009ma}, the second finding is still not understood.
For the puzzling In-In dip a suggestion has been made in \cite{Blaschke2011137}
where this feature could be reproduced within the generalized Matsui approach
by a subtle interplay of screening and absorption in the parametrization of 
the temperature dependence of the oscillator frequency. 
For the details see \cite{Carlosthesis:2012}. 
According to this picture, the dip reflects a nonmonotonous temperature 
behavior of the confining potential due to a resonance-like contribution 
in $\RE[\omega^2(T)]$.
This calculation is rather susceptible to changes in the absorptive part in 
the complex oscillator strength, basically 
as a consequence of the Dirac delta distribution for the initial $Q\bar{Q}$ 
state.

\begin{figure}[htb]
	\centering
\includegraphics[width=0.9\columnwidth]{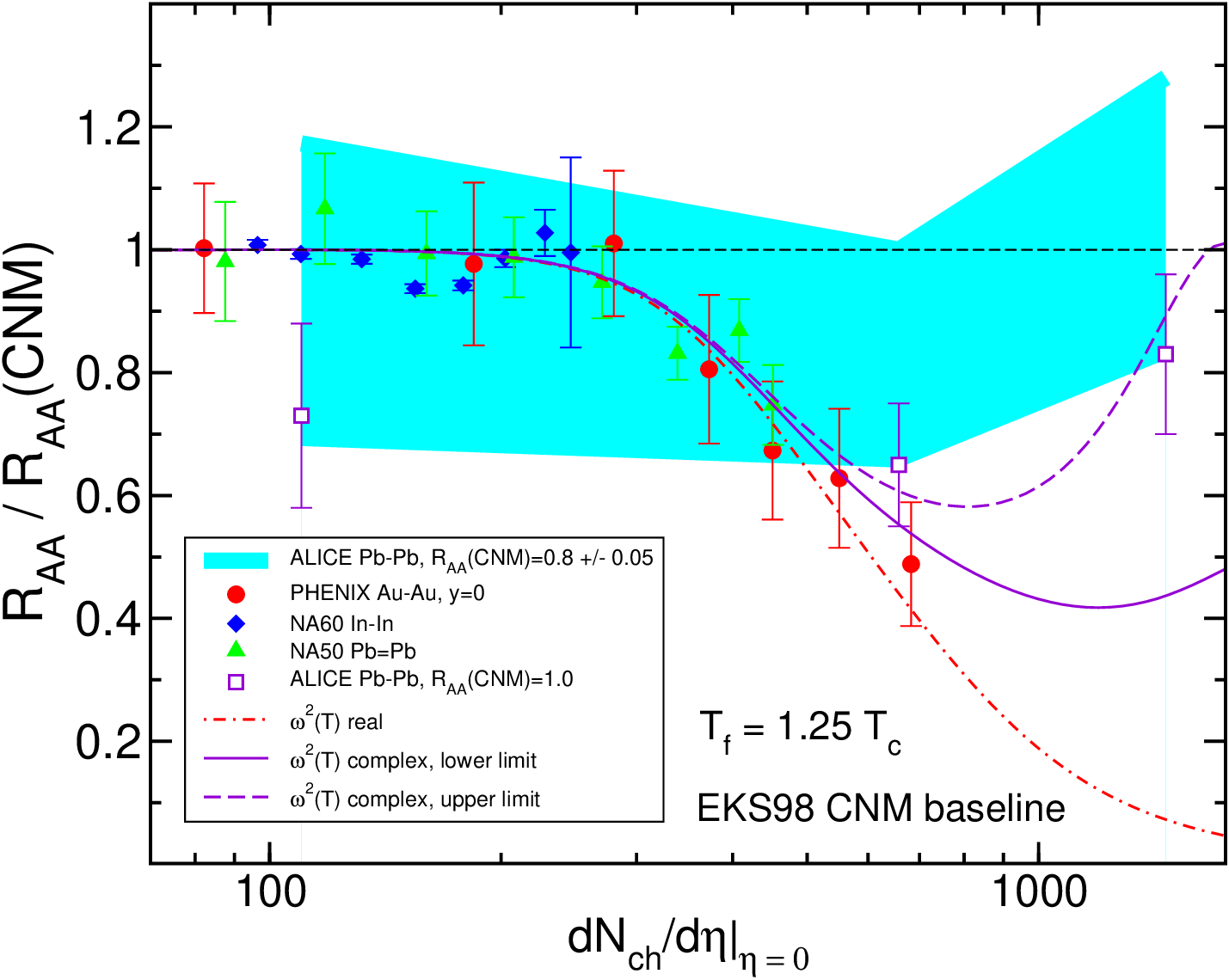}
\includegraphics[width=0.9\columnwidth]{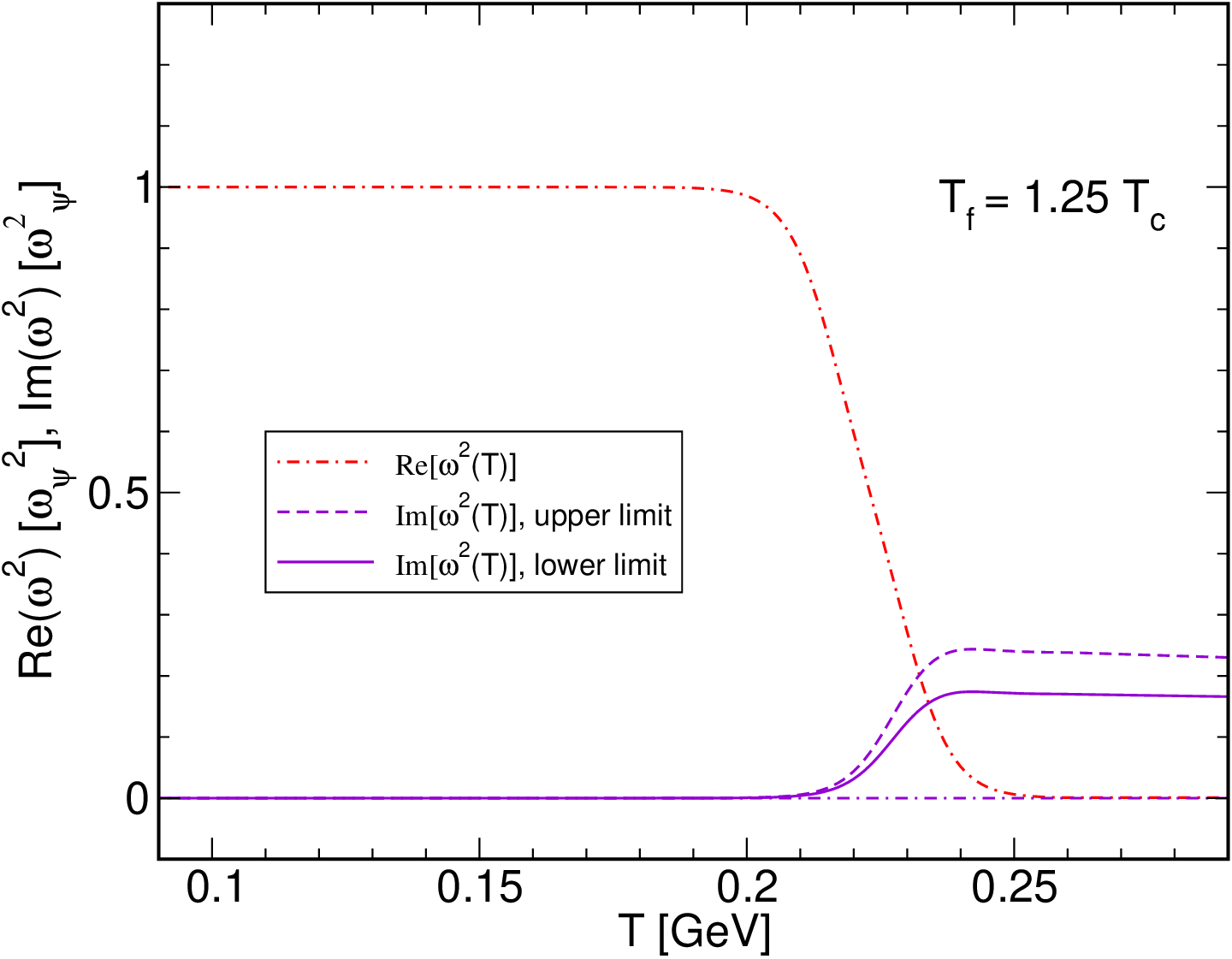}
	\caption{Predictions of the quantum mechanical THO model for 
charmonium suppression at CERN and RHIC. 
The LHC data are described by a purely imaginary part of the $Q\bar{Q}$ 
Hamiltonian.}
	\label{fig:Jpsi41}
\end{figure}

\begin{figure}[htb]
	\centering
\includegraphics[width=0.9\columnwidth]{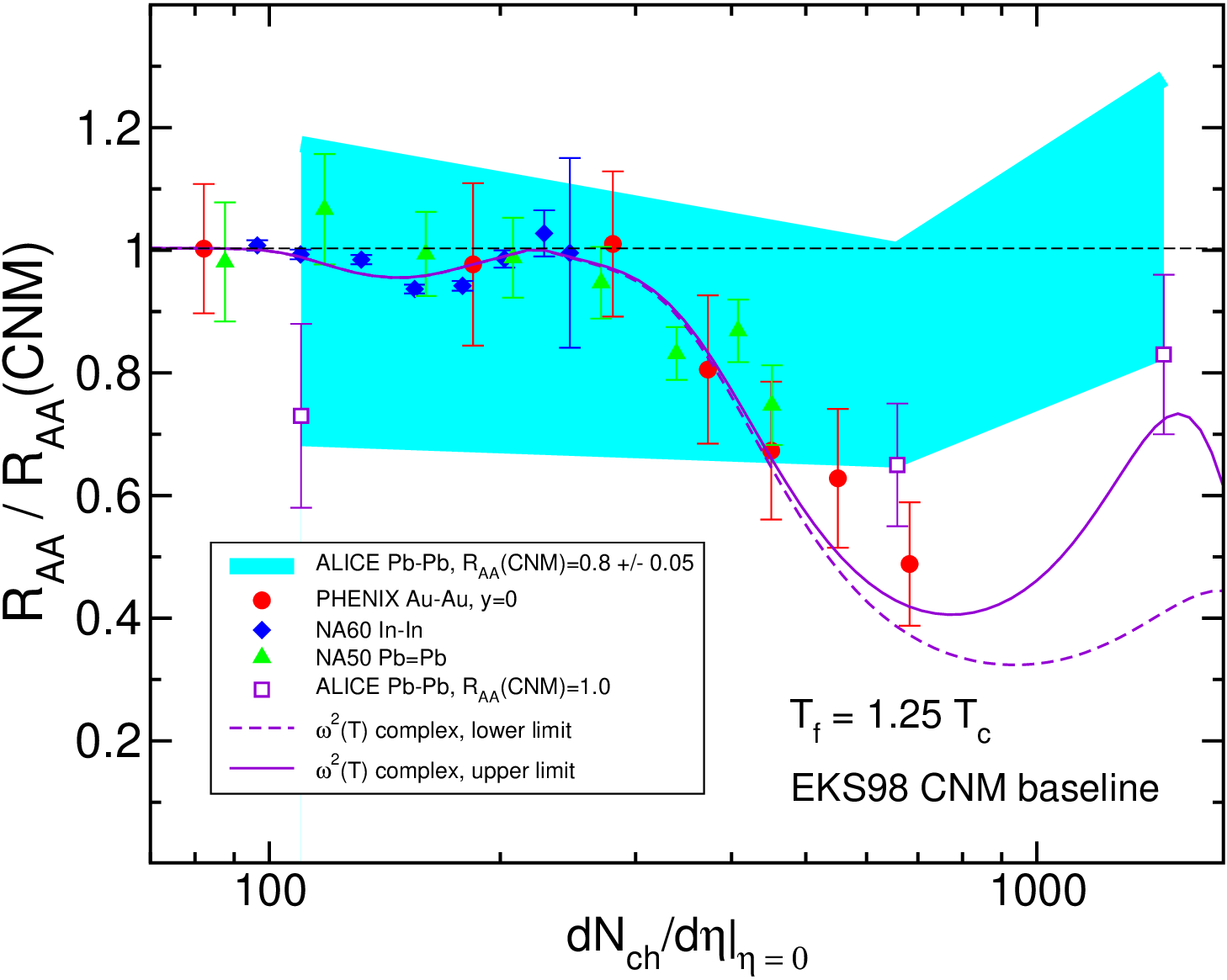}
\includegraphics[width=0.9\columnwidth]{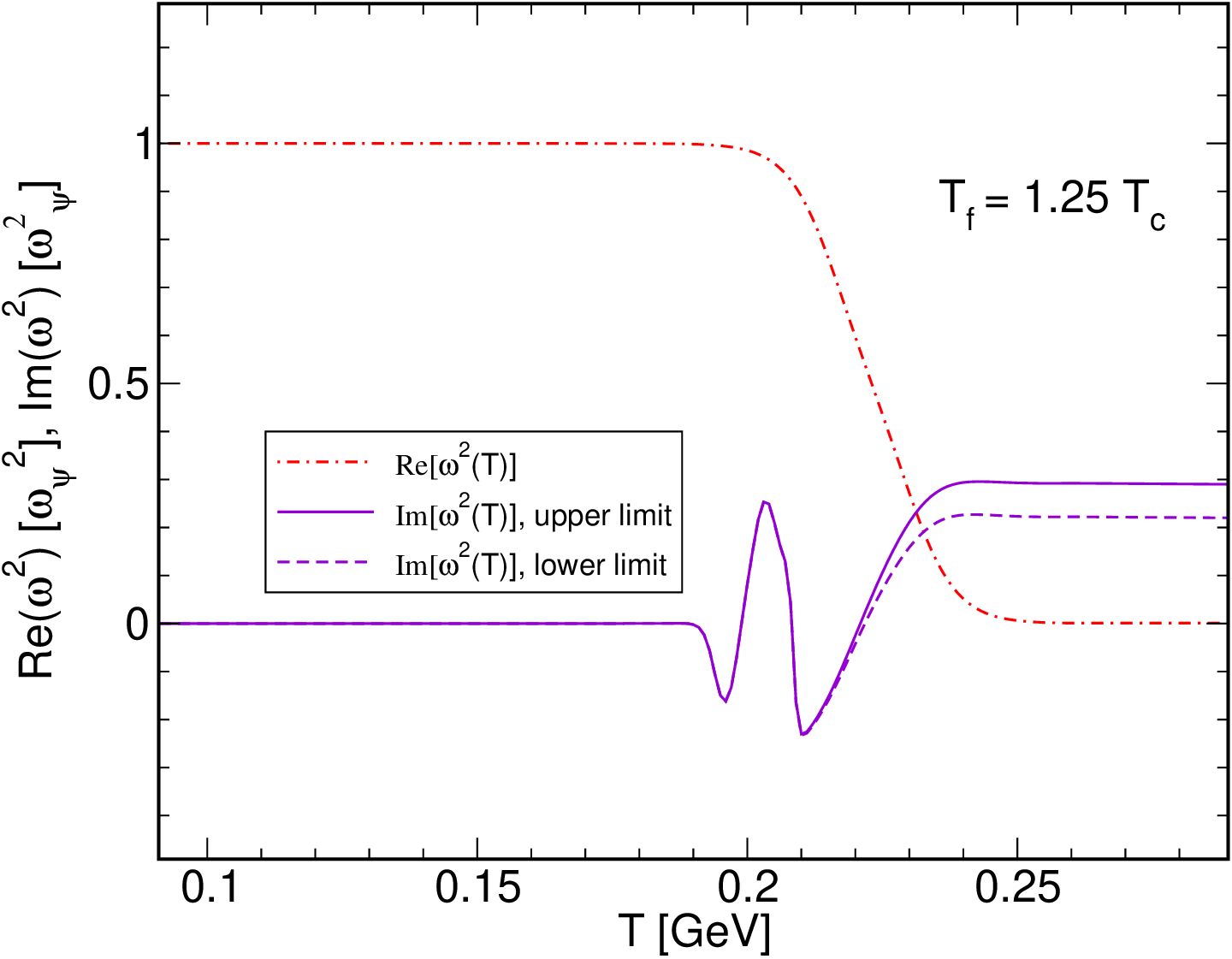}
	\caption{The same as Fig.~\ref{fig:Jpsi41}, but fitting the ``wiggle'' 
in the In-In data by varying the imaginary part of the frequency. 
The dip is mostly described by absorption and regeneration on the region from  
0.193~GeV to 0.21~GeV.
The screening is rather visible above $T_f=1.25~T_c$, as a consequence the 
production of charmonium decreases despite of the presence of the imaginary 
part. In order to exhibit a regeneration pattern in the domain of LHC beyond 
the RHIC data, 
the imaginary part of the squared oscillator frequency for temperatures above 
0.23~GeV shall be larger than the corresponding one given in 
Fig.~\ref{fig:Jpsi41}.}
	\label{fig:Jpsi42}
\end{figure}

\begin{figure}[htb]
	\centering
\includegraphics[width=0.9\columnwidth]{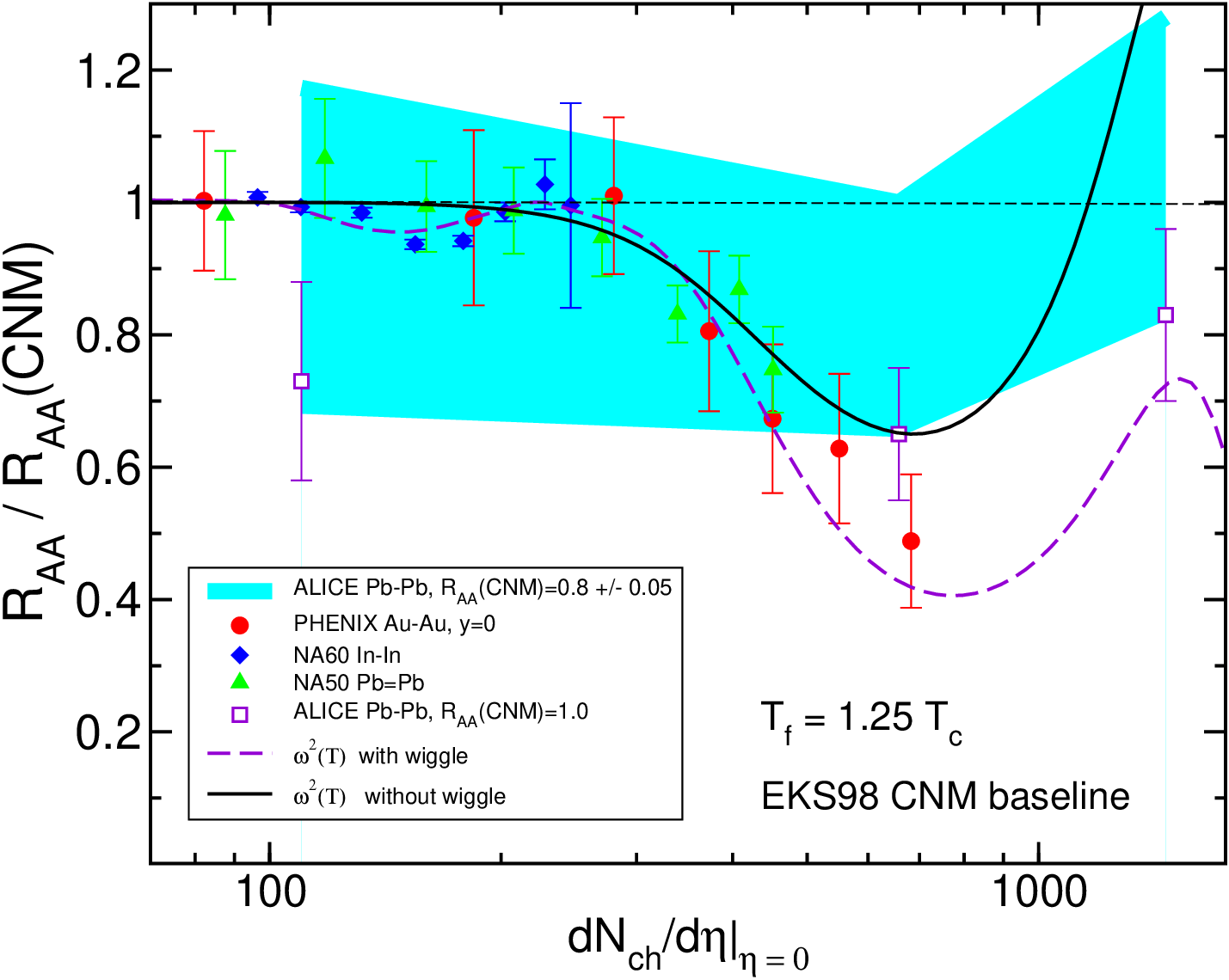}
	\includegraphics[width=0.9\columnwidth]{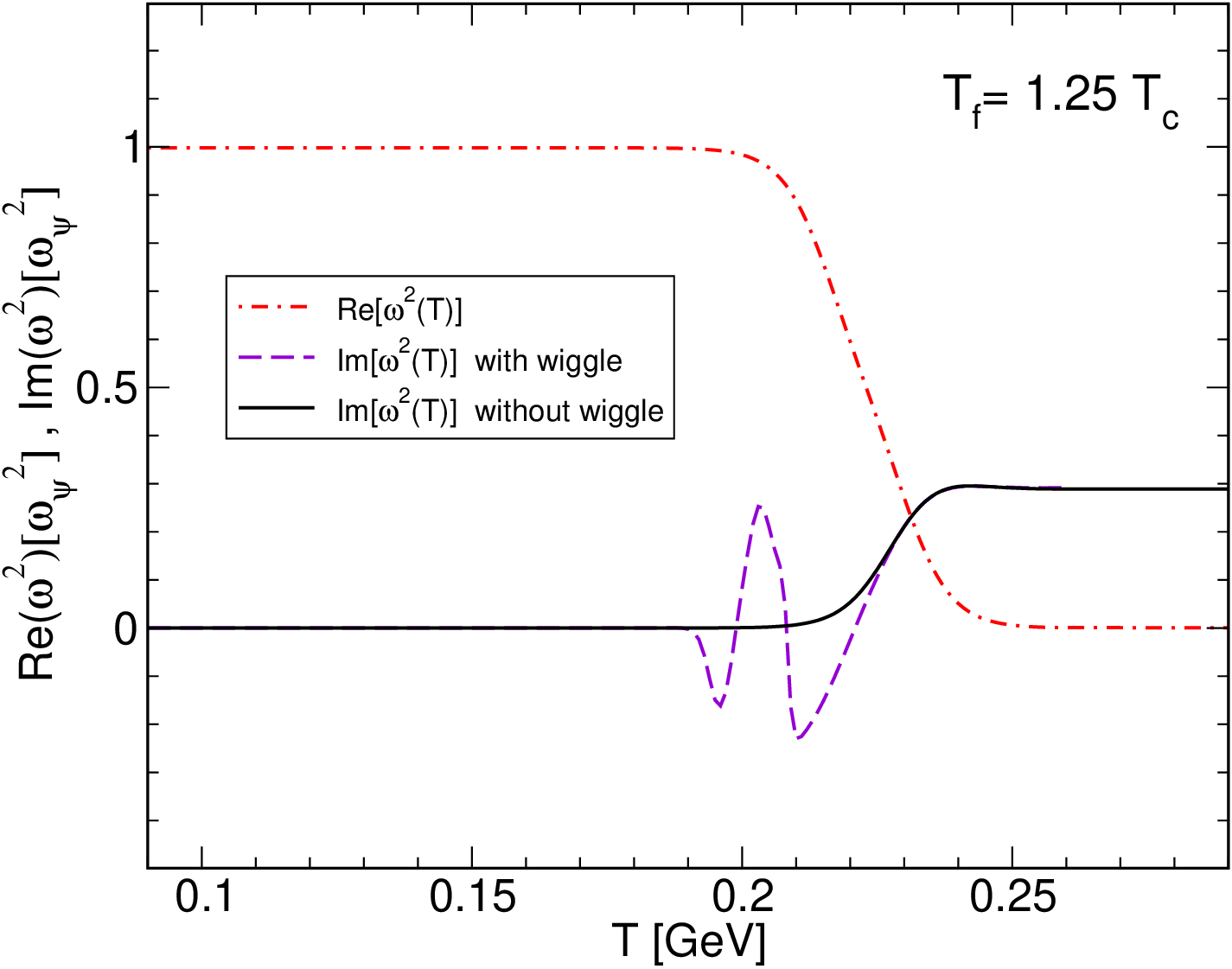}
	\caption{Influence of the wiggle on the charmonium suppression at
the LHC region. This region correspond to temperatures above 0.23 GeV see 
reference \cite{Carlosthesis:2012}.
The two LHC predictions are completely different above 
$d N_{\rm ch}/d\eta \sim 700$ although the complex oscillator strength above 
0.23 GeV is the same. The difference is due to the different values of 
$\IM [\omega^2(T_f)]$.}
	\label{fig:Jpsi43}
\end{figure}

In the present work we investigate the role of $\IM[\omega^2(T)]$ in 
reproducing the dip in In-In data, see Fig.~\ref{fig:Jpsi41} and 
Fig.~\ref{fig:Jpsi42}. 
The parametrization used here allows a description of screening, absorption 
and regeneration simultaneously.
Additionally, we employ for the initial $Q\bar{Q}$ state the gaussian shape 
given in  \cite{Carlosthesis:2012}.
For the LHC data on J/$\psi$ suppression we employ recent results of the 
ALICE collaboration \cite{Scomparin:2012vq}. For estimating the CNM effects 
which are not measured yet at LHC, we deduce an error band guided by 
theoretical estimates in Ref.~\cite{Vogt:2010aa} and shown as the cyan hatched
region in Figs.~\ref{fig:Jpsi41}-\ref{fig:Jpsi43}.

For an additional discussion of the behavior of the imaginary part of the 
$Q\bar{Q}$ potential shown in Fig.~\ref{fig:Jpsi42} we refer to a separate 
analysis of the $Q\bar{Q}$ plasma Hamiltonian 
\cite{Blaschke:2009uh} and the discussion of the role of
the X(3872) particle and the spectral function of $\rho$ meson in a quark 
meson plasma in this context \cite{Blaschke2011137,Carlosthesis:2012}.
It is worth noting that the wiggle pattern in the In-In data is located in the 
region of temperatures $1.2 \leq T/T_c < 1.5$ 
\cite{Blaschke2011137, Carlosthesis:2012}. 
The influence of this wiggle on the charmonium suppression in the LHC region 
with charged particle densities exceeding the ones reached at RHIC is explored 
in Fig.~\ref{fig:Jpsi43}. 
In this figure the two LHC predictions are completely different
above $d N_{\rm ch}/d\eta \sim 700$ although the complex oscillator strength 
is the same above 0.23 GeV. 
How can this wiggle affect the LHC predictions so strongly? 
One explanation can be seen in the survival probability (\ref{spsi}) which 
encodes the quantum time evolution of the quarkonium state. 
As we have described above, the time evolution of the complex oscillator 
frequency is implied by its assumed temperature dependence using the 
temperature evolution of the plasma $T(\tau)$ which follows from entropy 
conservation under Bjorken expansion  (\ref{entropy}) with the fit of the 
lattice QCD data for the entropy density (\ref{eos-fit}). 

The predictions for LHC are affected by the value of the oscillator 
frequency at $T_0(0)=T_f\sim 1.25~T_c$ where the complex frequency $\omega(0)$ 
enters the transition amplitude $U(\vec{r},\vec{r}_0,t)$ 
(see Eq.~(\ref{evolution})) through the condition 
$\rho(0)=1/\sqrt{\omega(0)}$ and is determined by the equality 
$\omega(0)=\omega(T_0(0))=\omega(T_f)$.

\section{Conclusion}

Recent studies of heavy quarkonia correlators and
spectral functions at finite temperature in lattice QCD and
systematic T-matrix approaches using QCD motivated finite-temperature
potentials support that heavy quarkonia dissociation shall occur in the
temperature range $1.2 \le T/T_c \le 1.5$ whereby the interplay of both
screening and absorption processes is important.
We have discussed these effects on the quantum mechanical evolution of
quarkonia states within a time-dependent harmonic oscillator model with
complex oscillator strength and compared the results for the survival 
probability with data for $R_{\rm AA}/R_{\rm AA}$(CNM) from SPS, RHIC
and LHC experiments.
Besides the traditional interpretation, with a threshold for the onset of
anomalous suppression by screening and dissociation kinetics at
$d N_{\rm ch}/d\eta \sim 300$, we suggest an alternative arising from the
attempt to model the dip of the suppression pattern of the rather precise NA60
data from In-In collisions at $d N_{\rm ch}/d\eta \sim 150 - 250$.
We suggest that this dip indicates the true threshold for the onset of
anomalous suppression due to the coupling of charmonium to the 
$\bar D^{*0}D^0$ channel with the recently discovered $X(3872)$ state.
Although some details have been worked out in 
\cite{Carlosx:2011cp,Carlosthesis:2012}, the theoretical basis for supporting 
this hypothesis has apparently been developed in plasma physics with the
concept of a plasma Hamiltonian for nonrelativistic bound states like heavy
quarkonia when they are immersed in a medium dominated by strong correlations
like bound states. 
In the context of the theory of strongly correlated plasmas the formation 
and dissociation of bound states can be systematically addressed within 
cluster expansion techniques. 
With the advent of the new quality of the data from the ALICE experiment at 
LHC which provide yields of hidden ($Q\bar{Q}=$ J/$\psi$, $\Upsilon$, ...) 
{\it and} open heavy flavor ($Q\bar{q}=$ D, B, ...) mesons, it may be 
possible for the first time to measure the quantity 
$n_{Q\bar{Q}}/n_{Q\bar{q}}n_{q\bar{Q}}$ which one can denote as heavy quark 
association degree in analogy to a similar quantity in plasma physics. 
Such a quantity is of advantage for the QGP diagnostics not only because it is
known for plasma diagnostics in stellar atmospheres, semiconductor plasmas and
nuclear matter, but in particular since it provides a new baseline for 
quantifying the J/$\psi$ suppression effect.
We will return to this issue in a separate work.

In the present work we have illustrated an alternative to the models of 
charmonium kinetics and statistical hadronization in the description of 
data for anomalous J/$\psi$ suppression that utilizes the quantum mechanical
formulation of quarkonia production with a complex Hamiltonian.  
To illustrate the method, a simple THO model with complex oscillator 
frequencies has been presented whereby the temperature dependence of the 
real and imaginary parts resulting from fitting heavy-ion collisions data 
of SPS, RHIC and LHC experiments on quarkonium production may allow to discuss
the interplay of screening, absorption and regeneration of heavy quarkonia
in these experiments. 
It is obvious that this tool may be further refined by using realistic 
temperature dependent complex potentials as motivated by lattice QCD studies.
The quantum mechanical approach can also be exploited, e.g., to study the 
relative abundances of different excited states (Saha equation aspect of 
plasma diagnostics) and to study provide insights about the role of resonances
at the open charm thresholds (most prominently the X(3872)) for the 
regeneration of J/$\psi$ in an analogy to the role of the Hoyle state in 
element synthesis \cite{Hoyle:1954}. 

\section*{Acknowledgments}
We would like to thank R.~Arnaldi, E.~Scomparin and M.~Leitch for providing
us with the data shown in Figs.~\ref{fig:Jpsi41} - \ref{fig:Jpsi43}. 
Comments by P.~Braun-\-Munzinger to an earlier version of this paper helped us
to improve the presentation and are gratefully acknowledged.
This work was supported by the National Science Centre (NCN)
within the ``Maestro" programme under contract DEC-2011/02/A/ST2/00306.

\end{document}